\documentclass[a4paper]{article}
\PassOptionsToPackage{bookmarks=false}{hyperref}
\usepackage{ISCSLP2021,subfigure,multirow,array,cite}
\usepackage[backref]{hyperref} 
\hypersetup{
colorlinks=true,
linkcolor=black,
citecolor=black,
urlcolor=black
}
\usepackage{url}
\title{Accent and Speaker Disentanglement in Many-to-many Voice Conversion}

\name{Zhichao Wang$^1$, Wenshuo Ge$^1$, Xiong Wang$^1$,Shan Yang$^1$,Wendong Gan$^2$,Haitao Chen$^2$,Hai Li$^2$, Lei Xie$^1$$^*$, Xiulin Li$^3$}

\address{
  $^1$Audio, Speech and Language Processing Group (ASLP@NPU), School of Computer Science,
  Northwestern Polytechnical University, Xi’an, China\\
  $^2$iQIYI Inc, China\\
  $^3$Databaker (Beijing) Technology Co., Ltd.}
\email{zcwang\_aslp@mail.nwpu.edu.cn,sdjngws@gmail.com,\{syang,xwang\}@nwpu-aslp.org,\{Wendong Gan,Haitao Chen,Hai Li\}@qiyi.com,lxie@nwpu.edu.cn,lixiulin@data-baker.com}

\begin{document}

\maketitle
\begin{abstract}
This paper proposes an interesting voice and accent joint conversion approach, which can convert an arbitrary source speaker's voice to a target speaker with non-native accent. This problem is challenging as each target speaker only has training data in native accent and we need to disentangle accent and speaker information in the conversion model training and re-combine them in the conversion stage. In our recognition-synthesis conversion framework, we manage to solve this problem by two proposed tricks. First, we use accent-dependent speech recognizers to obtain bottleneck features for different accented speakers. This aims to wipe out other factors beyond the linguistic information in the BN features for conversion model training. Second, we propose to use adversarial training to better disentangle the speaker and accent information in our encoder-decoder based conversion model. Specifically, we plug an auxiliary speaker classifier to the encoder, trained with an adversarial loss to wipe out speaker information from the encoder output. Experiments show that our approach is superior to the baseline. The proposed tricks are quite effective in improving accentedness and audio quality and speaker similarity are well maintained.

\end{abstract}

\renewcommand{\thefootnote}{\fnsymbol{footnote}}
\footnotetext{* Lei Xie is the corresponding author.}

\section{Introduction}

Voice conversion (VC) aims to modify speech from a source speaker to sound like that of a target speaker without changing the linguistic content. Accent conversion (AC) focuses on transforming non-native speech to sound as if the speaker had a native accent. VC transforms speaker identity while maintaining the linguistic information. By contrast, AC transforms accents while maintaining the speaker identity as well as the linguistic content. Early approaches on both directions involves frequency warping\cite{Sundermann2003,Erro2010,Godoy2012,Aryal2014}, exemplar methods\cite{Takashima2012,ZWU2014} and Gaussian mixture models (GMM)\cite{Stylianou1998,Toda2007}. With the fast development of deep learning (DL), deep neural network (DNN)\cite{Desai2009,Sun2015,hsu2017voice,Zhao2018,Zhang2019} based conversion has significantly improved the performance recently due to its strong feature learning and non-linear mapping ability.

According to the data can be used, conversion methods can be parallel and non-parallel. Parallel conversion needs parallel utterances of the same linguistic content uttered by source and target speakers (for VC) or source and target accents by the same speaker (for AC). By contrast, non-parallel conversion is more practical but also more challenging, which does not necessarily need parallel data with the same linguistic content but need to disentangle the linguistic and speaker/accent information.
 
Targeting to non-parallel conversion, there are two major frameworks for both voice conversion and accent conversion: phonetic posteriorgram (PPG)\cite{Zhao2018,7552917,Zhao2019} based two-stage approach and end-to-end neural approach\cite{Tian2019,qian2019autovc,chou2019oneshot}. The former recognition-synthesis framework first adopts an individually trained speech recognition system to transform source speaker (or accent) speech into an intermediate feature representation, either PPG or bottleneck features (BN) that extracted from a hidden layer of an ASR acoustic model, and then a DNN-based conversion model is trained to map PPG or BN into acoustic features of the target speaker (or accent). Inspired by recent success of sequence-to-sequence (seq2seq) modeling in speech recognition and synthesis,  end-to-end neural VC\cite{8936924, Liu2020} approach has also emerged with an encoder-decoder framework where the encoder and the decoder function as recognition and conversion modules respectively.

This paper focuses on an interesting and more challenging conversion task -- voice and accent joint conversion, where the source speaker's voice can be converted to the target speaker with the desired accent and unchanged linguistic content. We suppose a more practical many-to-many non-parallel conversion scenario, where we have data from a set of target speakers each has a unique native accent. So the challenging problem is how to disentangle linguistic, speaker, and accent information simultaneously in the same conversion framework, and re-combine them flexibly during the conversion stage as $a_m$ is not his/her native accent. More specifically, we have non-parallel data from speaker-accent combination $\{s_i$,$a_j\}$, where speaker $s_i$ only can speak with native accent $a_j$, and for an arbitrary source speaker, we aim to convert his or her voice to any desired speaker-accent combination, e.g., $\{s_n$,$a_m\}$, where speaker $s_n$ does not have data with accent $a_m$ for system building.

To disentangle linguistic information and the other two factors, we choose a recognition-synthesis framework. Specifically, we adopt a well-trained speech recognizer to first transform the source speaker's voice to bottleneck features. Inspired by \cite{Liu2020,Zhou2019}, we use accent-dependent speech recognizers to obtain BN features for different accented speakers. This aims to further disentangle the linguistic information from other factors including accents in the BN features for conversion model training. And then we use a Tacotron-like conversion model with separate accent ID and speaker ID to control the target speaker and accent in conversion. Different from Tacotron \cite{wang2017tacotron}, we use the duration from the ASR model for frame expansion instead of attention-based soft alignment. Inspired by~\cite{8936924,dat}, we propose to use adversarial training to better disentangle the speaker and accent information. Specifically, we plug an auxiliary speaker classifier to the encoder, trained with an adversarial loss to wipe out speaker information from the encoder output. Experiments have shown that our approach is superior to the baseline, and the proposed tricks are quite effective in improving accentedness. 

With our voice and accent joint many-to-many conversion framework as well as the accent and speaker disentanglement approach, we are able to convert an arbitrary source speaker's voice to a target speaker with desired accent. We manage to disentangle the speaker and the associated accent in a speaker set during the conversion model training stage and re-combine them as desired during the conversion stage.


\section{Related work}

Although there are many early approaches, the \textit{recognition-synthesis} approaches, either hybrid or end-to-end, are the mainstream solution to non-parallel voice conversion and accent conversion. Sun et al.\cite{7552917} first proposed the phonetic posteriorgram based VC method where the phonetic posteriorgram serves as an intermediate linguistic representation as input to a speech synthesis model trained for the target speaker. Later, accent conversion has also successfully adopted this framework \cite{Zhao2018,Zhao2019}. With the fast development of deep learning, especially the seq2seq mapping, VC and AC have both embraced the end-to-end framework \cite{Tian2019,qian2019autovc,chou2019oneshot,8936924,Liu2020} with a unified neural network model for `recognition' and `synthesis'.

As discussed in Section 1, a key problem in voice and accent conversion is information disentanglement, such as disentangling linguistic and speaker information. A similar problem has been investigated in seq2seq based text to speech for several years \cite{cong2020data,yang2020adversarial}. In \cite{cong2020data}, noise and speaker are disentangled via domain adversarial training (DAT) \cite{dat} for multi-speaker TTS in the wild. Similar idea has been expanded to more generalized noise case \cite{yang2020adversarial} and voice cloning based on few shots and one shot \cite{cong2020data}. Built upon the encoder-decoder framework, a recent voice conversion study adopted adversarial training strategy to wipe out speaker information from the linguistic representations, which brought the best performance in voice conversion challenge \cite{8936924}.

Our work is mostly inspired by the cross-lingual conversion based on modularized neural network~\cite{Zhou2019}, accent conversion using accented ASR to learn accent-agnostic linguistic representations~\cite{Liu2020} and the above adversarial learning approaches~\cite{8936924,yang2020adversarial,cong2020data}. In this work, we particularly focus on the voice and accent joint many-to-many conversion task, where the source voice from an arbitrary speaker can be converted to a target speaker (from a target set) with specified accent and more challengingly, the target speaker does not have the training data for this specified accent.

\section{Proposed approach}
\label{sec:proposedapproach}

In this paper, we choose standard Mandarin (M) and Tianjin-accented Mandarin (T) as two target accents. Tianjin dialect is a distinctly recognizable Mandarin accent. While there are four major tones in Chinese Mandarin, the Tianjin dialect seems to prefer three of the tones over the first one (the high and level one). This preference makes the accent sound sloppy. There are three target speakers $\{s_1,s_2,s_3\}$, in which $\{s_1,s_2\}$ speak in standard Mandarin while the other $\{s_3\}$ speaks in Tianjin accent. Our aim is to convert any speaker's voice to the three target speakers with designated accent (M or T), e.g., $s_2$ with T. Note that our method can be applied to other accents as well. For simplicity, we study the above two accents in the paper.

\vspace{-3pt}
\subsection{Overview and model architecture}
\label{ssec:overview}

\begin{figure}[ht]
  \centering
  \vspace{-10pt}
  \includegraphics[width=1\linewidth]{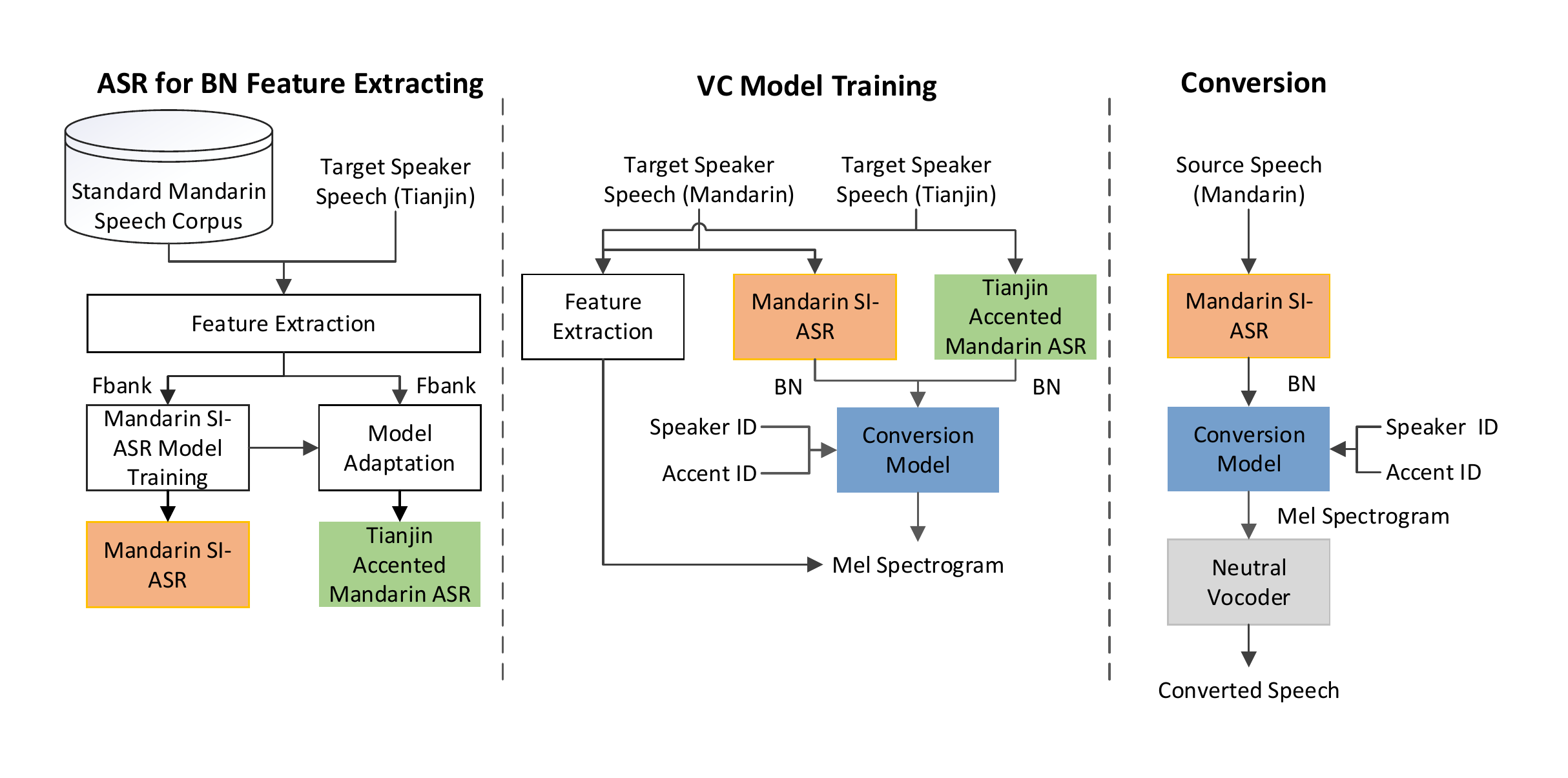}
  \vspace{-10pt}
	\caption{Schematic diagram of the proposed approach.}
  \label{fig:training_procedure}
  \vspace{-10pt}
\end{figure}

As shown in Fig.~\ref{fig:training_procedure}, our approach is composed of three stages. As we follow a typical recognition-synthesis approach, the first stage is to train the ASR acoustic model to extract bottleneck features which will be subsequently used as the input to the voice conversion model. Our VC model aims to predict mel-spectrogram from BN features. In order to model multiple target speakers and accents, we use speaker ID and accent ID as auxiliary signal along with the training data to the conversion model. Then in the conversion stage, these auxiliary labels are used as control to designate the desired speaker and accent. However, each speaker only has his/her native accented speech for training and speaker and accented are intermined. To sovle this problem, we propose several tricks to make the control more effective, for example, letting the standard Mandarin speaker speaks in Tianjin accent or vise versa, which will be introduced in the following subsections.

\begin{figure}[ht]
  \centering
  \vspace{-10pt}
	\includegraphics[width=1\linewidth]{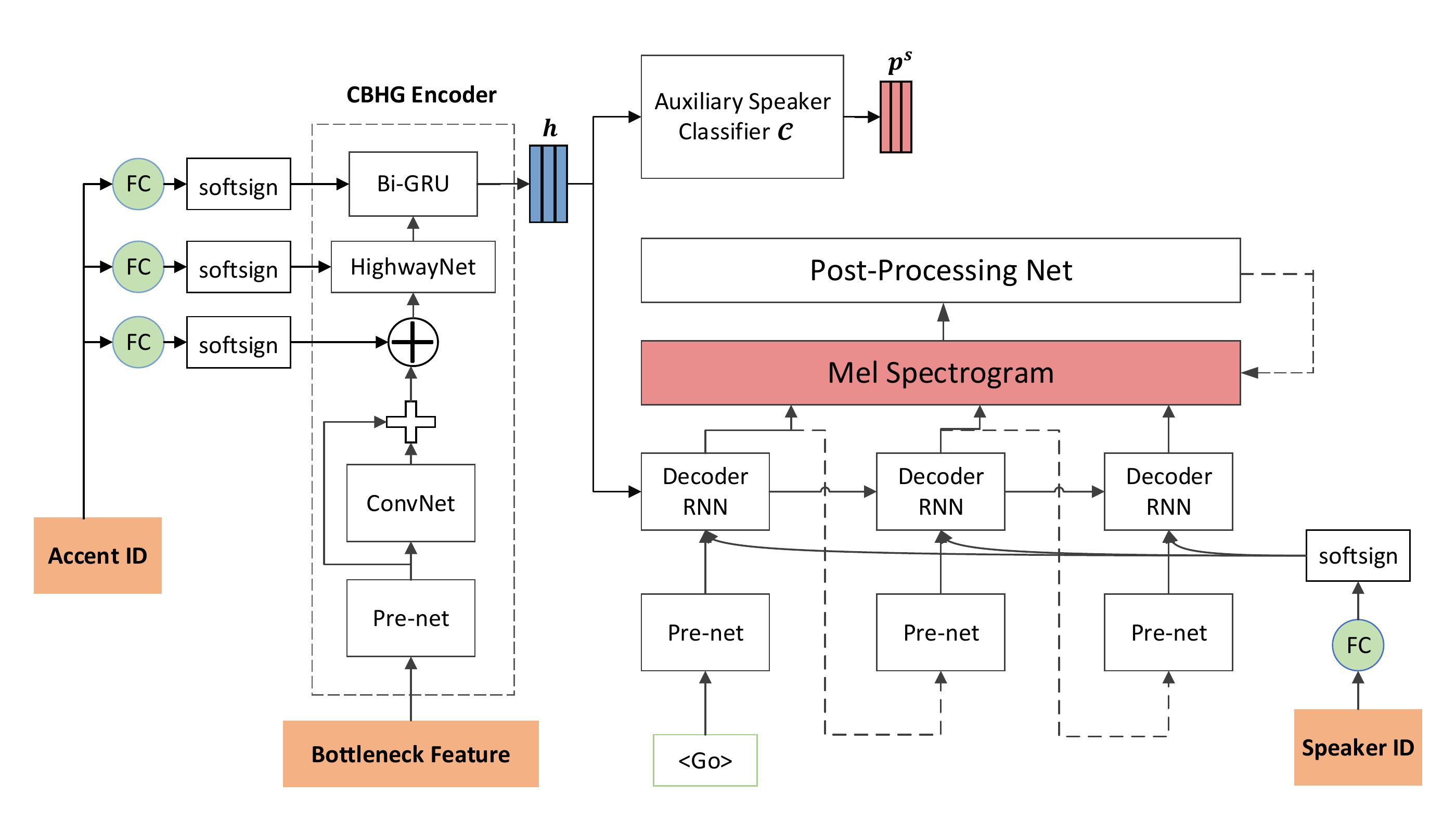}
	\caption{The network architecture of the conversion model.}
  \label{fig:model_structure}
  \vspace{-10pt}
\end{figure}

As shown in Fig.~\ref{fig:model_structure}, the conversion model follows a typical encoder-decoder architecture. Similar to Tacotron~\cite{wang2017tacotron}, we use the CBHG module as the encoder, and meanwhile the decoder adopts an auto-regressive module consisted of prenet, decoder RNN and postnet. Different from Tacotron, we use the duration from the ASR model for frame expansion instead of attention-based soft alignment. To represent and control the speaker identity and accent, we separately add accent embedding into the encoder and speaker embedding into the decoder~\cite{Cao2019}. To wipe out speaker-related information in the encoder output, we add an auxiliary speaker classifier after the encoder and adversarial training strategy is adopted, which will be introduced in detail in Section 3.3.

\subsection{Accent-dependent ASR model}

As mentioned in Section 1, the bottleneck features extracted by ASR may leak nonlinguistic information (such as accent-related information) to the voice conversion model. Ideally we desire the BN features only contain pure linguistic information. This is particularly important to our case as we designate the target speaker with a desired accent. To this end, we use accent-dependent ASR acoustic model to obtain the BN features for different accented target speakers during the conversion model training. We first train an ASR system using a large standard Mandarin corpus, denoted as Mandarin SI-ASR in Figure 1. Then we adapt this model using the data from the target speaker with Tianjin accent and the resulting ASR is denoted as Tianjin-accented Mandarin ASR in Figure 1. The two ASR systems are subsequently used to extract the corresponding BN features for different accented speakers during VC model training.

\subsection{Adversarial Training Against Speaker Classification}

\begin{figure}[ht]
  \vspace{-15pt}
	\centering
  \includegraphics[width=0.8\linewidth]{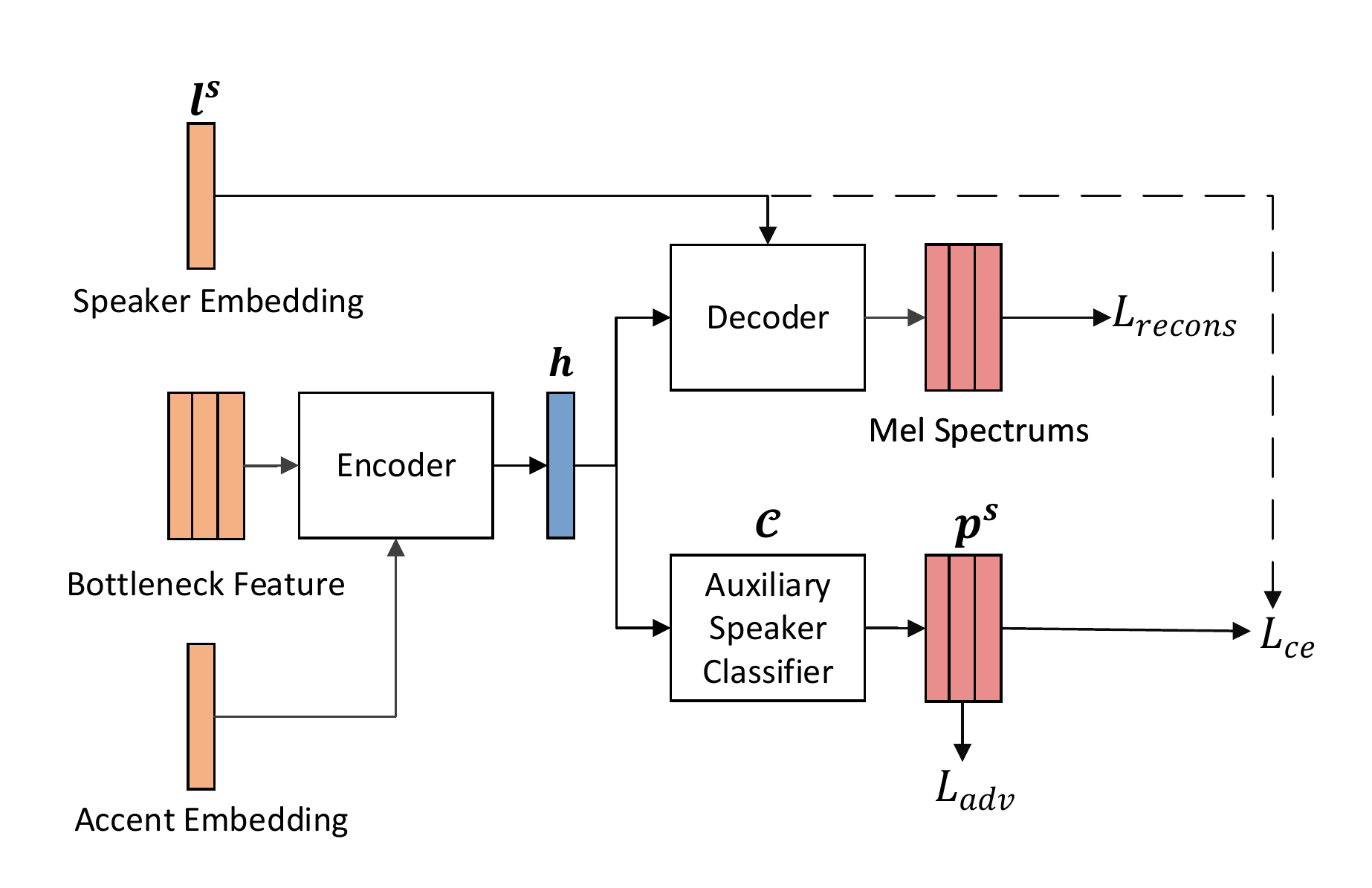}
  \vspace{-10pt}
	\caption{The schematic diagram of model losses.}
  \label{fig:speaker_confusion}
  \vspace{-15pt}
\end{figure}

In order to better disentangle speaker identity, as shown in Fig.~\ref{fig:speaker_confusion}, we use an auxiliary speaker classifier with an adversarial training strategy (ADV) to obtain speaker-independent text representations. 

The auxiliary speaker classifier $\mathcal{{C}}$ receives the hidden text representations $\mathbf{h}$ as inputs and predicts the speaker identity:
\vspace{-2.5pt}
\begin{equation}
  {p}^{s}=\mathcal{C}(\mathbf{h})
\end{equation}
where ${p}^s$ is the predicted probability for speaker $s$, and the auxiliary classifier is trained with cross-entropy loss $L_{ce} = \text{CE}({p}^s, {l}^s)$ to predict the speaker identity from the $\mathbf{h}$ where $l^s$ is the one-hot speaker label. 

In order to disentangle the speaker information in $\mathbf{h}$, we tend to confuse the classifier to not be able to distinguish the speaker classes. To achieve this goal, we propose to force the speaker logistics of classifier to obey a uniform distribution. In details, we adopt an adversarial loss to the encoder as
\vspace{-2pt}
\begin{equation}
  L_{adv} = ||{p}^s - {e}||_2^2
  \label{equation:loss2}
\end{equation}
where ${p}^s$ is the logistic of speaker classifier, $e=[1/N, \ldots, 1/N]$ represents a uniform distribution while ${N}$ is the number of speakers. During estimating the speaker classifier, the parameters of encoder are frozen. In this way, minimizing the above $L_{adv}$ will assign equal probability to each possible speaker given $\mathbf{h}$, which means that the classifier cannot distinguish the speaker identities.

With the extra speaker classification network as discriminator $D$ and other parts of VC model as generator $G$, the final adversarial objective function becomes

\vspace{-2pt}
\begin{equation}
   \left\{ \begin{array}{l}
    Loss_G=L_{recons}+\beta L_{adv}\\
    \\
    Loss_D=L_{ce}
      \end{array} \right.
  \label{equation:lossall}
\end{equation}
where $L_{recons}$ denotes the mel-spectrogram reconstruction loss, and $\beta$ is the weight for the classification loss. $G$ and $D$ are trained alternately. When training $G$, the parameters of $D$ are frozen and vise versa.

\section{Experiments}
\label{sub:experiments}

\subsection{Dataset and Experimental Setup}
\label{datasetandfeatureextraction}

We conduct voice and accent joint conversion experiments on three target speakers, where $s_1$ and $s_2$ are male and female standard Mandarin speaker respectively and $s_3$ is a female Tianjin-accented speaker. Each speaker has about 5000 sentences with 5-hours of speech. For conversion test, we select another 10 Mandarin speakers (5 females and 5 males) as the source speakers and covert 10 utterances from each source speaker to the three target speakers with two different accents (M and T). All speech utternaces are 16kHz and 16bit. We use 80-dim mel-spectrogram as features computed in 50ms frame length and 12.5ms frame shift. Our Mandarin SI-ASR acoustic system is a TDNN-F model implemented using the Kaldi toolkit~\cite{Povey} with the 30,000 hours of Mandarin training data. We use the 256-dim bottleneck features as the linguistic representation which is extracted from the last fully-connected layer before softmax. Then we use the speech  data from the Tianjin-accented speaker ($s_3$) to fine-tune the above TDNN-F model which is subsequently used to extract BN features for this speaker in VC model training. We use a `universal' melLPCnet vocoder, which is similar to LPCnet~\cite{valin2018lpcnet} but uses mel-spectrogram as input to generate waveform. The vocoder is trained using 200 hours of clean speech from 21 speakers.

We implement 3 systems for ablation study. They are:
\begin{itemize}
	\item \textbf{BL: }the baseline system which uses the Mandarin SI-ASR for BN feature extraction during VC model training and conversion. The conversion model follows the structure in Figure 2 but does not have the auxiliary speaker classifier and ADV.
	\item \textbf{P1: }the system uses accent-dependent ASR for BN feature extraction during VC model training. The conversion model follows the structure in Figure 2 but does not have the auxiliary speaker classifier and ADV.
	\item \textbf{P2: }the system uses accent-dependent ASR for BN feature extraction during VC model training. The conversion model follows the structure in Figure 2 with auxiliary speaker classifier and ADV.
\end{itemize} 

In the conversion model training stage, conversion model is trained for 90 epochs using batch size of 32. We use Adam optimizer~\cite{kingma2014adam} with learning rate decay, which starts from 0.001 and decays every 15 epochs in decay rate 0.7. For $L_{recons}$, we use a simple MSE loss for both outputs before and after post-net. The two losses have equal weights. Due to the strong correlation between speaker and accent, in order to accelerate model convergence, we do use accent representation after 10 epochs of the model training. Specifically, in P2, the conversion model and auxiliary speaker classifier are set to alternate every 5 epochs of training, and $\beta$ is set to 0.3.

\begin{figure}[ht]
  \centering
  \subfigure[P1 (w/o ADV)]{
  \begin{minipage}[t]{0.48\linewidth}
  \centering
  \includegraphics[width=1.3in]{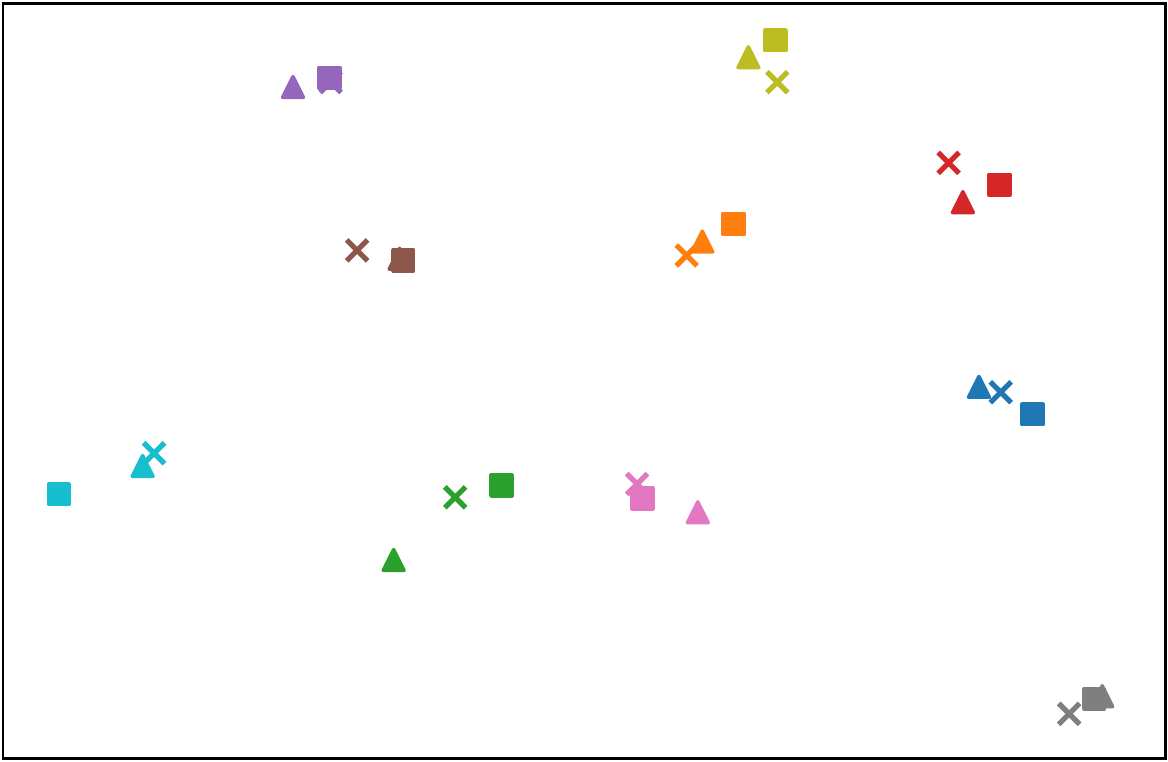}
  \end{minipage}%
  }%
  \subfigure[P2 (w/ ADV)]{
  \begin{minipage}[t]{0.48\linewidth}
  \centering
  \includegraphics[width=1.3in]{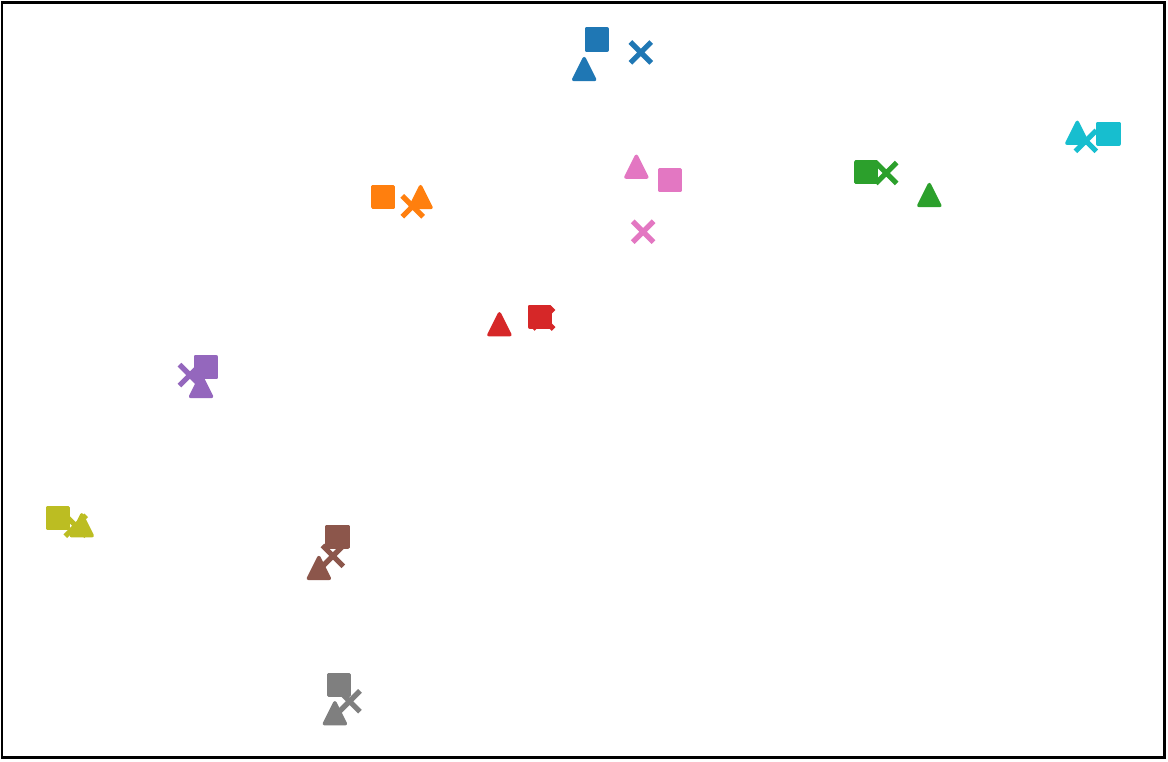}
  \end{minipage}%
  }%
  \vspace{-10pt}
  \caption{Encoder output visualization using t-SNE. Each color represents a sentence while different-shaped symbols represent different speakers.}
\label{fig:lin_t-sne}
\vspace{-15pt}
\end{figure}

\subsection{Visualization of Encoder Output}

To demonstrate the ability to remove speaker-related information for encoder output using ADV, we visualize the hidden representations by t-SNE~\cite{VanDerMaaten2008} for system P1 and P2. Ten parallel utterances from 3 standard Mandarin source speakers are selected and sent into the encoder for visualization. The encoder output ${h}^r$ is averaged along the time axis to obtain a single embedding vector for each utterance. Then the embedding vectors are projected into a 2-dimensional space by t-SNE. Comparing (a) and (b) in Fig.~\ref{fig:lin_t-sne}, we can see that after ADV, the utterances with the same linguistic content (the same color) but from different speakers (the different shape) are grouped more closely. This indicates that the adversarial learning is effective to remove speaker-related information and help encoder to generate speaker-invariant output.

\subsection{Speaker Similarity}

\begin{figure}[ht]
  \vspace{-10pt}
  \centering
  \begin{minipage}[t]{0.33\linewidth}
  \centering
  \includegraphics[width=1in]{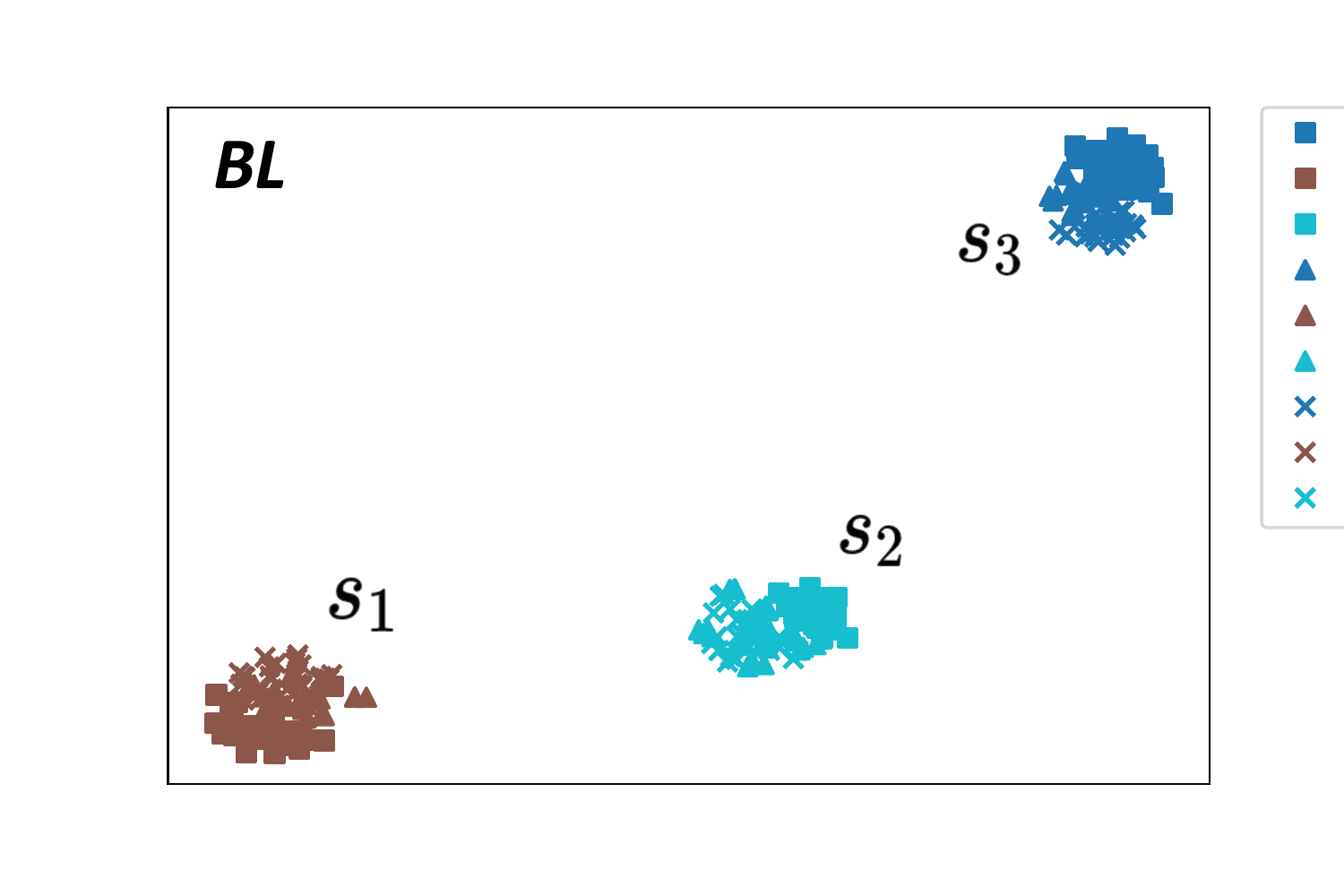}
  \end{minipage}%
  \begin{minipage}[t]{0.33\linewidth}
  \centering
  \includegraphics[width=1in]{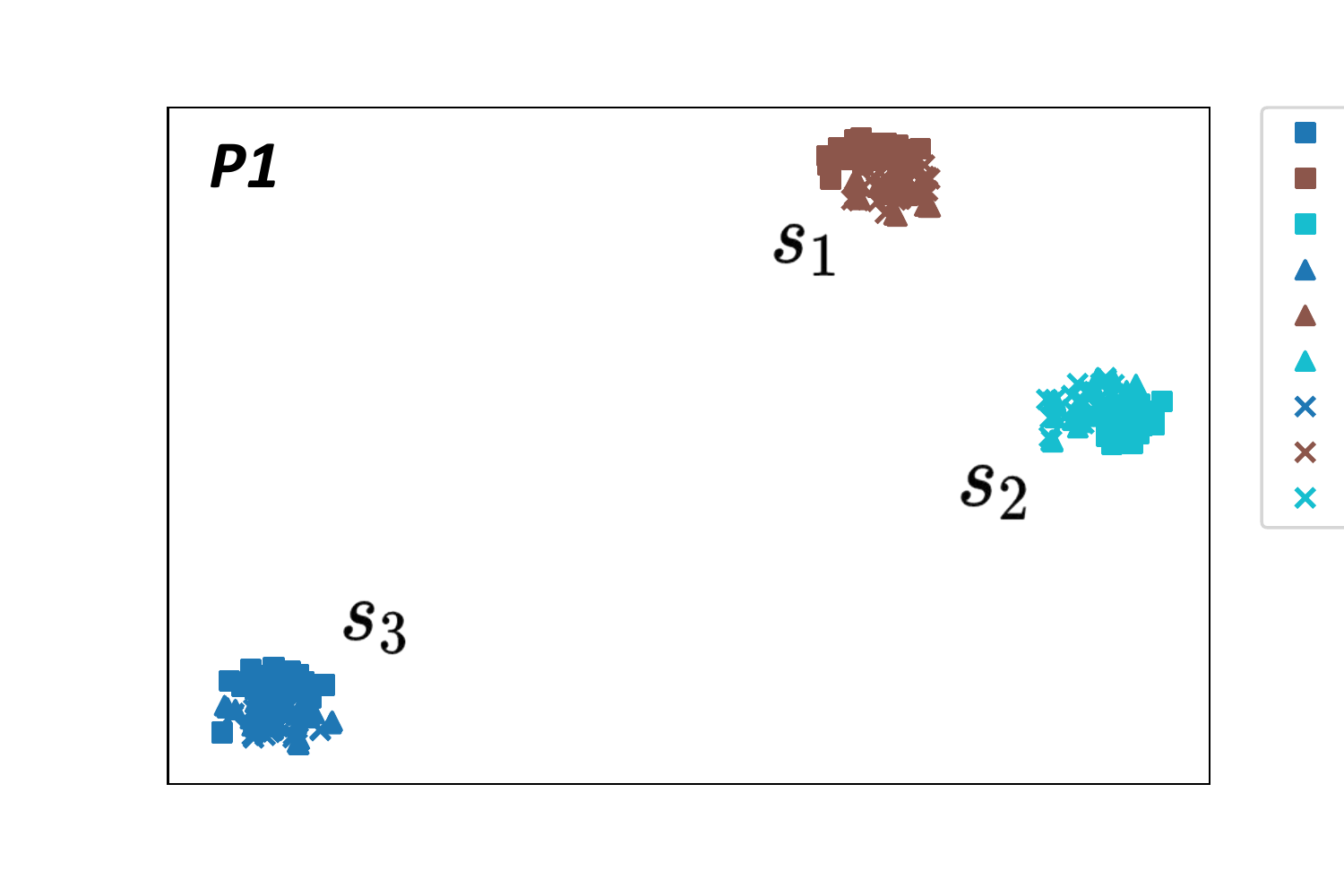}
  \end{minipage}%
  \begin{minipage}[t]{0.33\linewidth}
  \centering
  \includegraphics[width=1in]{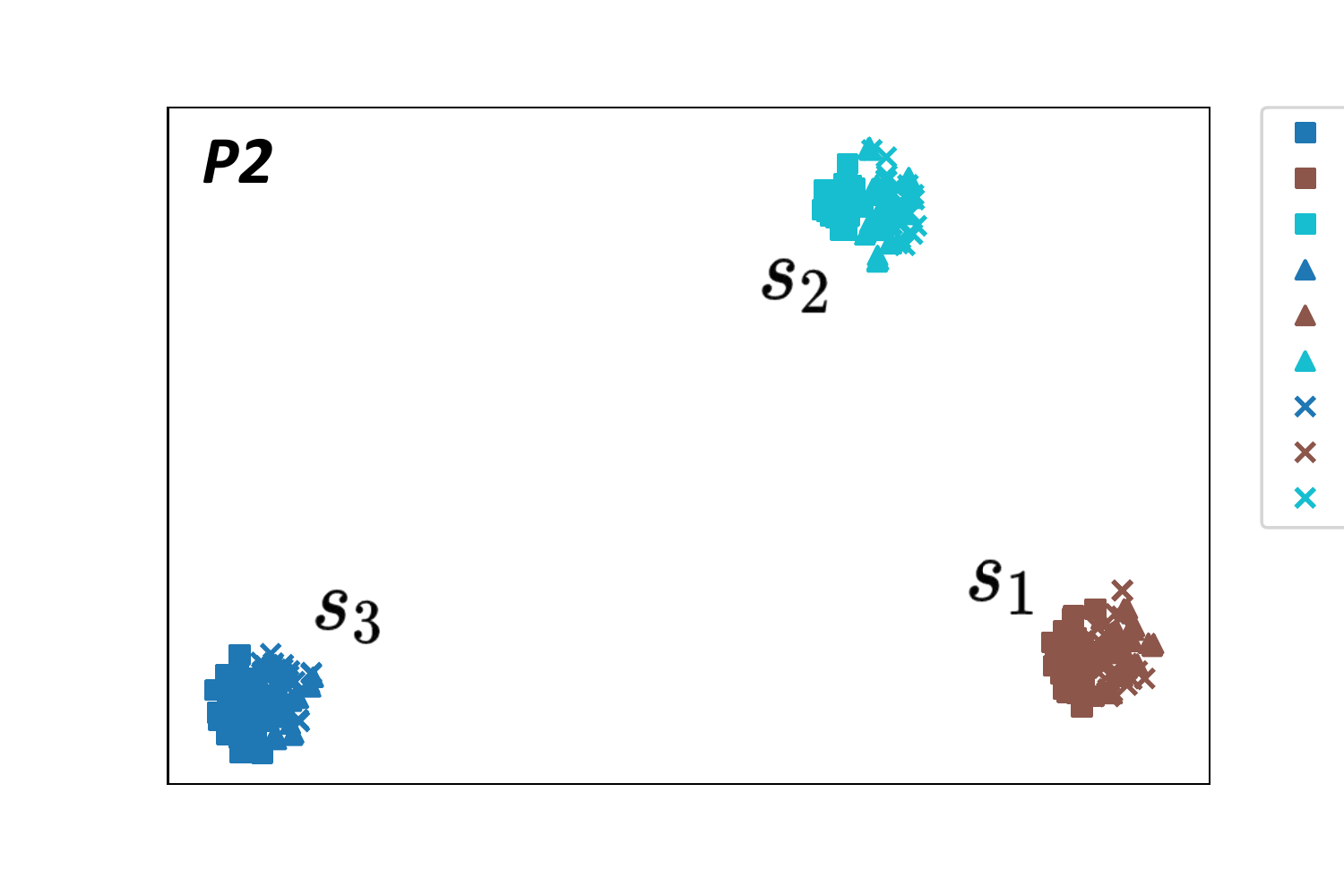}
\end{minipage}%

\caption{X-vector visualization using t-SNE. Different color represents different speaker. Different shaped symbols represent the converted accent speech (M and T) and the original speech.}
\label{fig:t-sne}
\vspace{-15pt}
\end{figure}

To verify the speaker similarity between the converted speech and the original target speech, we perform speaker visualization based on x-vector. We train an x-vector~\cite{Snyder2018} extractor using 2000 hours of Mandarin speech from 16125 speakers. The model architecture is Resnet32 optimized with additive margin softmax loss~\cite{Wang_2018} and the embedding size is 1024. For each model, we randomly select 30 utterances from the 10 source speakers and convert them to 3 target speakers each with 2 accents. Then we extract the x-vectors for the 180 ($30\times 3 \times 2$) converted samples and another 30 utterances from each target speaker in the training set. We visualize these totally 270 x-vectors by t-SNE in Figure~\ref{fig:t-sne}. We can see that no matter what system used (B1 and P1,2), all samples are grouped into three clusters representing the three target speakers. This unveils that the output speech samples from the three conversion models, including those converted samples with non-native new accent, have successfully preserved the speaker similarity of the target speakers. 

\subsection{Subjective Evaluation}

Two listening tests are conducted to evaluate the performance of the three systems: a mean opinion score (MOS) test of audio quality and an AB test on accentedness. We randomly select 20 utterances from the converted samples for each model for the listening tests, which contain all possible conversion pairs $\{s_i, a_j\}$. A group of 12 Chinese speakers has participated in the tests. The converted samples can be grouped into two subsets according to the accent ID  -- M and T, each subset contains samples with 3 target speakers (speaker ID). We highly recommend the listeners to listen to our samples\footnote{Samples can be found in \href{https://kerwinchao.github.io/AccentVoicejointConversion.github.io/}{\url{https://kerwinchao.github.io/AccentVoicejointConversion.github.io/}}}.

$\textbf{Audio Quality}.$ In the MOS test, listeners are asked to rate the quality of the converted speech on a 5-point scale. Audios converted from the three systems are randomly shuffled before presenting to listeners. Each group of audio corresponds to the same text content. The MOS results in Table~\ref{tab:mos} show that the three systems achieves similar MOS values on the synthetic samples when the target accent is set to M. When the target accent is set to T, the quality MOS becomes lower, but the P2 system achieves the best quality MOS as compared with other systems. Note that the baseline system (BL) cannot generate samples properly with Tianjin accent (T). Even we set accent ID to T, the converted samples are still more similar to standard Mandarin accent. Hence the BL system does not have samples for MOS test.


\begin{table}[ht]
  \vspace{-5pt}
  \caption{MOS results with 95\% confidence interval.}
  \vspace{-5pt}
  \label{tab:mos}
  \centering 
  \scriptsize
  \renewcommand\arraystretch{1.5}
  \begin{tabular}{m{0.2cm}<{\centering}|m{1.8cm}|m{1.4cm}<{\centering}m{1.3cm}<{\centering}m{1.2cm}<{\centering}}
  \hline
  \textbf{ID} &\textbf{Model}                 &\textbf{Accent ID:M} &\textbf{Accent ID:T} & \textbf{Total}      \\ \hline
  BL &Baseline                        & \textbf{3.93$\pm$0.157}                               & -                              &-          \\ \hline
  P1 &\ + Accent-D ASR                      & 3.90$\pm$0.152                               & 3.67$\pm$0.173                              & 3.78$\pm$0.117          \\ \hline
  P2 &\ \ \ + ADV & 3.91$\pm$0.153                      & \textbf{3.79$\pm$0.183}                     & \textbf{3.85$\pm$0.120} \\ \hline
  \end{tabular}
  \vspace{-7pt}
  \end{table}

\begin{figure}[ht]
  \vspace{-15pt}
    \centering
    \includegraphics[width=1.0\linewidth]{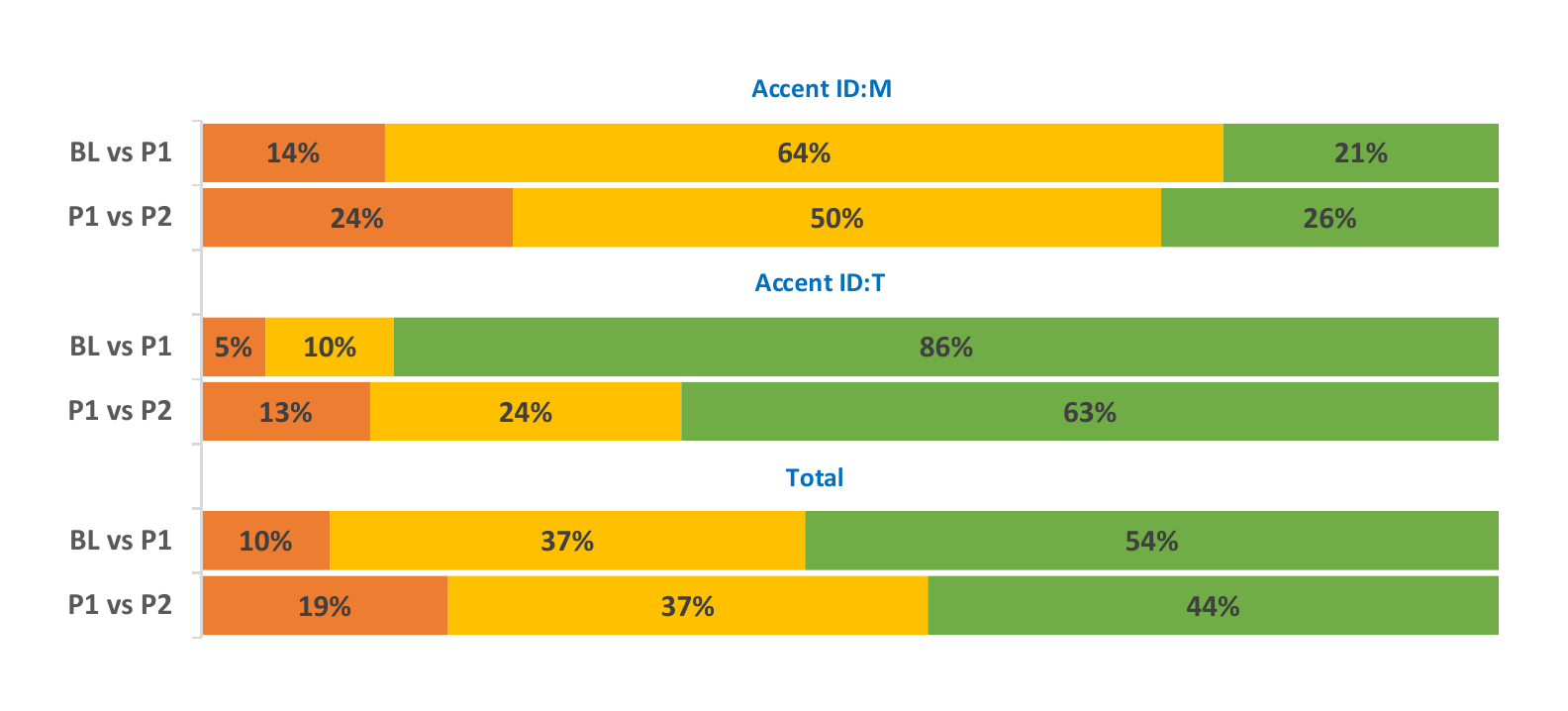}
    \vspace{-20pt}
  \caption{Accentedness preference test results (A:NP:B).}
  \vspace{-15pt}
  \label{fig:ABTEST}
\end{figure}

\textbf{Accentedness}. In the AB test on accentedness, paired speech samples with the same textual content are presented and the listeners are asked to choose samples that are more similar to the target accent. The results are shown in Fig.~\ref{fig:ABTEST}. When the target accent is Mandarin (M), the proposed tricks are effective with a little bit more preference. When the target accent is Tianjin (T), the proposed tricks are more effective with much more preference. Specifically, the use of accent-dependent ASR (P1) has very high preference over the baseline system and the difference between P2 and P1 are also significant. These results show that accent-dependent ASR and adversarial training stratergy are both benifical to the perceived accentedness. Finally the best accentedness is achieved by the P2 system which adopts the two tricks at the same time.

\section{Conclusion}

We propose a voice and accent joint many-to-many conversion framework as well as several accent and speaker disentanglement methods. With these contributions, we can convert an arbitrary source speaker's voice to a target speaker with desired accent. 
This task is challenging as each target speaker only has training data in native accent and we need to disentangle accent and speaker information in the conversion model training and re-combine them in the conversion stage.
We manage to solve this difficult problem by using accent-dependent ASR for bottleneck feature extraction and adversarial learning for disentanglement of speaker and accent. We plan to expand our approach to more speaker-accent pairs and test it in other languages such as English.



\bibliographystyle{IEEEtran}

\bibliography{mybib}

\end{document}